%% file: main.tex
\documentclass[twocolumn,showpacs,aps,prl,superscriptaddress,floatfix]{revtex4}

\usepackage{graphicx}
\usepackage{dcolumn}
\usepackage{amsmath}
\usepackage{epsfig}

\input pubboard/babarsym.tex

\newcommand{\BABARPubYear}    {07}
\newcommand{\BABARPubNumber}  {057}
\newcommand{\SLACPubNumber} {13040}

\def\Bf         {{\it BF}\xspace}
\def\bll {\ensuremath {\Bz \to l^{+} l^{\prime -}}}

\def\bee {\ensuremath {\Bz \to e^{+} e^{-}}}
\def\bmm {\ensuremath {\Bz \to \mu^{+} \mu^{-}}}
\def\bem {\ensuremath {\Bz \to e^{\pm} \mu^{\mp}}}
\def\de        {\ensuremath {\Delta E}}

\begin{document}

\title{\begin{flushleft}
       \mbox{\normalsize {\babar}-PUB-\BABARPubYear/\BABARPubNumber} \\
       \mbox{\normalsize SLAC-PUB-\SLACPubNumber}
       \end{flushleft}
       \vskip 20pt
       Search for decays of {\boldmath $B^0$} mesons into {\boldmath $e^+e^-$} , \\
{\boldmath $\mu^+\mu^-$ }, and {\boldmath $e^{\pm}\mu^{\mp}$} final states}

\input pubboard/authors_jul2007.tex

\date{\today}

\begin{abstract}
We present a search for the decays \bee, \bmm and \bem using data
collected with the \babar\ detector at the PEP-II $\epem$ 
collider at SLAC. Using a dataset corresponding to $384 \times 10^6$
\BB pairs, we do not find evidence of any of the three decay modes.
We obtain upper limit on the branching fractions, at $90\%$ confidence
level, of ${\cal B}(\bee)< 11.3 \times 10^{-8}$, ${\cal B}(\bmm) <
5.2\times 10^{-8}$, and ${\cal B}(\bem) < 9.2\times 10^{-8}$.
\end{abstract}


\maketitle

The standard model (SM) of particle physics does not allow flavor
changing neutral currents at tree-level, and decays of this kind are
predicted to have very small branching fractions. This makes rare
decays particularly interesting for the detection of possible new
physics (NP) beyond the SM, such as supersymmetry~\cite{susy} (SUSY):
loop contributions from heavy partners of the SM particles predicted
in these models might induce, for certain decay modes, branching
fractions significantly larger than the values predicted by the SM.

\begin{figure}[t]
\begin{center}
\vskip 0.2cm
\hskip-4.5cm\includegraphics[width=3.3cm]{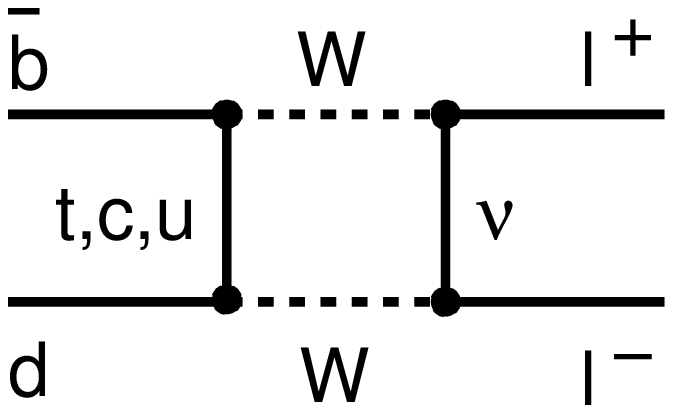}
\vskip -1.9cm
\hskip4.5cm\includegraphics[width=3.3cm]{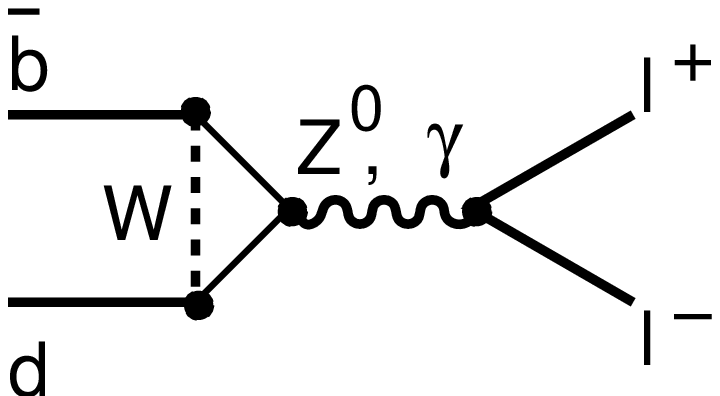}
\end{center}
\caption {Representative Feynman diagrams for $\Bz \to l^{+} l^{-}$ in the Standard
Model. }
\label{fig:bllfeyn}
\end{figure}

The leptonic decays $\bll$ (where $l^{+} l^{\prime -}$ stands for
$e^+e^-$, $\mu^+\mu^-$ or $e^{\pm}\mu^{\mp}$; charge conjugation is
implied throughout) are particularly interesting among rare decays,
since a prediction of the decay rate in the context of the SM can be
obtained with a relatively small error, due to the limited impact of
long-distance hadronic corrections~\cite{BRSM}.  In the SM, $\Bz \to
l^{+} l^-$ decays proceed through diagrams such as those shown in
Fig.~\ref{fig:bllfeyn}.  These contributions are highly suppressed
since they involve a $b \to d$ transition and require an internal
quark annihilation within the $B$ meson.  The decays are also helicity
suppressed by factors of $(m_{\ell}/m_B)^{2}$, where $m_\ell$ is the
mass of the lepton and $m_B$ the mass of the $B$ meson.

In addition, $B^0$ decays to leptons of two different flavors violate
lepton flavor conservation, so they are forbidden in the SM.  This
feature provides a handle to discriminate among different NP
models~\cite{burasemu}.

The \bll\ decays are sensitive to NP also in a large set of models
with Minimal Flavor Violation~\cite{MFV} (MFV), in which the NP
Lagrangian is flavor blind at the typical mass scale of new heavy
states, with reduced effects on flavor physics at the \B mass
scale~\cite{UUT}.  In the context of MFV models, NP corrections to
\bll\ are characterized by interesting correlations with other rare
decays for a particular choice of some fundamental parameters (as in
the case of small~\cite{bobeth} or large~\cite{gino} $\tan\beta$ in
SUSY models with MFV).  A precise determination of the decay rate of
$\bll$ would allow different NP scenarios to be disentangled.

\begin{table}[t]
\begin{center}
\caption{The expected branching fractions in the Standard
Model~\cite{BRSM} and the available upper limits (UL) at $90\%$ C.L.}
\vspace{0.1in}
\begin{small}
\begin{tabular}{lccc} \hline\hline\noalign{\vskip1pt}
Decay mode      & $B^0 \to e^+e^-$    & $B^0 \to \mu^+\mu^-$ & $B^0 \to e^\pm \mu^\mp$\\
\hline\noalign{\vskip1pt}
SM prediction   & $1.9\times10^{-15}$ & $8.0\times10^{-11}$  &           $0$            \\
\babar~\cite{babar} & $6.1 \times 10^{-8}$ & $8.3 \times 10^{-8}$ & $18 \times 10^{-8}$  \\
Belle~\cite{belle03} & $1.9 \times 10^{-7}$ & $1.6 \times 10^{-7}$ & $1.7 \times 10^{-7}$ \\
CDF~\cite{previousCDF}     &           -          & $2.3 \times 10^{-8}$ &         -     \\
CLEO~\cite{cleo}   & $8.3 \times 10^{-7}$ & $6.1 \times 10^{-7}$ & $15 \times 10^{-7}$  \\
\hline\hline\noalign{\vskip1pt}
\end{tabular}
\end{small}
\label{tab:expectedBR}
\end{center}
\end{table}

As shown in Table~\ref{tab:expectedBR}, the present experimental limits
on \bll\ are several orders of magnitude larger than SM expectations.
Nevertheless, improved experimental bounds will restrict the
allowed parameter space of several NP models.

The search for the $\Bz\to\tau^+\tau^-$ decay has been presented in a
previous paper~\cite{ref:tautau}.

In this paper, we present a search for \bll\ decay using data
collected with the \babar\ detector~\cite{babarnim} at the \pep2\
\epem storage ring at SLAC. The collider is operated at the \FourS
resonance with asymmetric beam energies, producing a boost
($\beta\gamma \approx 0.56$) of the \FourS along the collision axis.

The dataset used consists of $384 \times 10^6$ \BB pairs accumulated
at the \FourS resonance (``on-resonance''), equivalent to an
integrated luminosity of $347~\invfb$, and $37~\invfb$ accumulated at
a center-of-mass (CM) energy about $40\mev$ below the \FourS resonance
(``off-resonance''). The latter sample is used to characterize
background contributions not originating from $B$ decays.

Hadronic two body decays of $B$ mesons such as $B^0\to\pipi$ and
$B^0\to K^{\pm}\pi^{\mp}$ have the same event topology as the leptonic
ones and are therefore the main source of background from $B$ decays.
We use Monte Carlo (MC) simulations~\cite{geant} of $\bee$, $\bmm$,
$\bem$ decays (signal) and $B^0\to\pipi$ and $B^0\to K^{\pm}\pi^{\mp}$
(background) of approximately $3\times 10^5$ events each to optimize
event selection criteria and to estimate efficiencies.

Charged particles are detected and their momenta measured by the
combination of a silicon vertex tracker, consisting of five layers of
double-sided silicon detectors, and a 40-layer central drift chamber,
both operating in the 1.5-T magnetic field of a solenoid.  The
tracking system covers 92\% of the solid angle in the CM
frame. Identification of charged hadrons is provided by the average
energy loss (d$E$/d$x$) in the tracking devices and by an
internally-reflecting ring-imaging Cherenkov detector. For lepton
identification, we also use the energy deposit in the electromagnetic
calorimeter consisting of 6580 CsI(Tl) crystals and the pattern of
hits in resistive plate chambers (partially upgraded to limited
streamer tubes for a subset of the data used in this analysis)
interleaved with the passive material comprising the solenoid magnetic
flux return.

We reconstruct $\Bz$ meson candidates from two oppositely charged
tracks originating from a common vertex.  Signal events are
characterized by two kinematic quantities:
\begin{eqnarray}
\mes&\equiv& \sqrt{(s/2+{\bf p_0}\cdot{\bf p_B})^2/E_0^2-p_B^2} 
\label{eq1} \\
\de &\equiv& E_{B}^* - \sqrt{s}/2 ,
\end{eqnarray}
where $\sqrt{s}/2$ is the beam energy in the CM frame, the subscripts
$0$ and $B$ refer to the initial \FourS and to the $B$ candidate in
the laboratory frame, respectively, and the asterisk denotes the
\FourS rest frame.  In Eq.~(\ref{eq1}), the variable $s$ is used as
opposed to $E_B^*$ because $s$ is known with much greater precision
and the resulting correlation between $\mes$ and $\de$ is nearly zero.
For correctly reconstructed $B^0$ mesons, \mes peaks at the mass of
the $B^0$ meson with RMS of about 2.5 \mevcc, and $\Delta E$ peaks at
zero with RMS of about 25 \mev.  We require $|\DeltaE|< 150$ MeV and
$\mes>5.2$ GeV/c$^2$.

Since we use the pion mass hypothesis for the reconstruction of
tracks, the distribution of $\de$ peaks near zero for the $\pipi$ and
leptonic modes and at -50 MeV for $K^{\pm}\pi^{\mp}$.  The mass
hypothesis does not affect the distribution of $\mes$.

Energy loss by electrons due to final state radiation or
Bremsstrahlung in detector material leads to tails in the $\DeltaE$
and \mes distributions, in particular for the $\bee$ decay mode.  We
partially correct for this effect by adding the momentum of a photon
emitted at a small angle from the track to the electron momentum.

We apply stringent requirements on particle identification (PID) to
reduce the contamination from misidentified hadrons and leptons. In
this way, we retain $\sim 93\%$ ($\sim 73\%$) of the electrons
(muons), with a mis-identification rate for pions of less than $\sim
0.1\%$ ($\sim 3\%$).

According to the information provided by the PID, we separate our
dataset into three samples, $2e$, $2\mu$, and $1\mu 1e$, containing
events with two electrons, two muons and one muon and one electron,
respectively. The rest of the dataset ($h^{+}h^{\prime -}$) comprises
two oppositely charged hadrons and is used to characterize background
contamination to the three signal samples.

Based on MC simulations, we expect negligible cross feed of events
between the leptonic and hadronic data samples.

Contamination from non-resonant \qqbar ($q=u,d,s,c$) and $\tau^+
\tau^-$ production is reduced by exploiting their different event
topology with respect of that of the signal events. In particular, we
examine the distribution of final-state particles in the rest frame of
the \FourS candidate, in which the fragmentation of a \BB pair
(non-resonant event) produces an isotropic (jet-like) angular
distribution of the particles.

Non-\BB events are rejected by requiring the cosine of the sphericity
angle~\cite{sphericity} to be $|\cos\theta_S| < 0.8$, and the second
normalized Fox-Wolfram moment~\cite{foxwolf} to be $R_2 < 0.95$. In
addition, we use a Fisher discriminant~\cite{fisher} ($\cal{F}$) in
the maximum likelihood (ML) fit to separate the residual background
from signal events.  $\cal{F}$ is constructed from the CM momentum
$p_i$ and angle $\theta_i$ of each particle $i$ in the rest of the
event (ROE) with respect to the thrust axis~\cite{thrust} of the $B$
candidate.
\begin{equation}
{\cal F} \equiv 0.5319 - 0.6023L_0 + 1.2698L_2,
\label{Eq:fisher}
\end{equation}
where $L_0 \equiv \sum_i^{\rm ROE}p_i$ and $L_2 \equiv \sum_i^{\rm
ROE} {p_i \cos^2{\theta_i}}$.  The coefficients of the linear
combination have been optimized on samples of signal and background
simulated events. Since the variable $\cal{F}$ depends only on the
ROE, we use the same coefficients for the three leptonic decays in the
ML fit.

The background from other \BB events is found to be negligible after
applying the PID requirements.  Backgrounds originating from QED
events (electrons and muons coming from $\epem$ interactions) are
rejected by requiring more than four charged tracks in the event.

To ensure the quality of the measurement of the Cherenkov angle
$\theta_c$, we require more than five detected Cherenkov photons and
$\theta_c>0$.  For pion or lepton candidates, in order to reject
protons, we require $\theta_c$ to be within $4 \sigma$ of the value
expected for pions. For kaon candidates, we require $\theta_c$ to be
within $4 \sigma$ of the expected value for kaons.

Applying the criteria described above, we select 67 events in the $2e$
sample, 56 in the $2\mu$ sample, 86 in the $1\mu 1e$ sample and
$\approx 94\times 10^3$ in the $h^{+}h^{\prime -}$ sample.

\begin{figure*}[htbp]
\includegraphics[width=4.5cm]{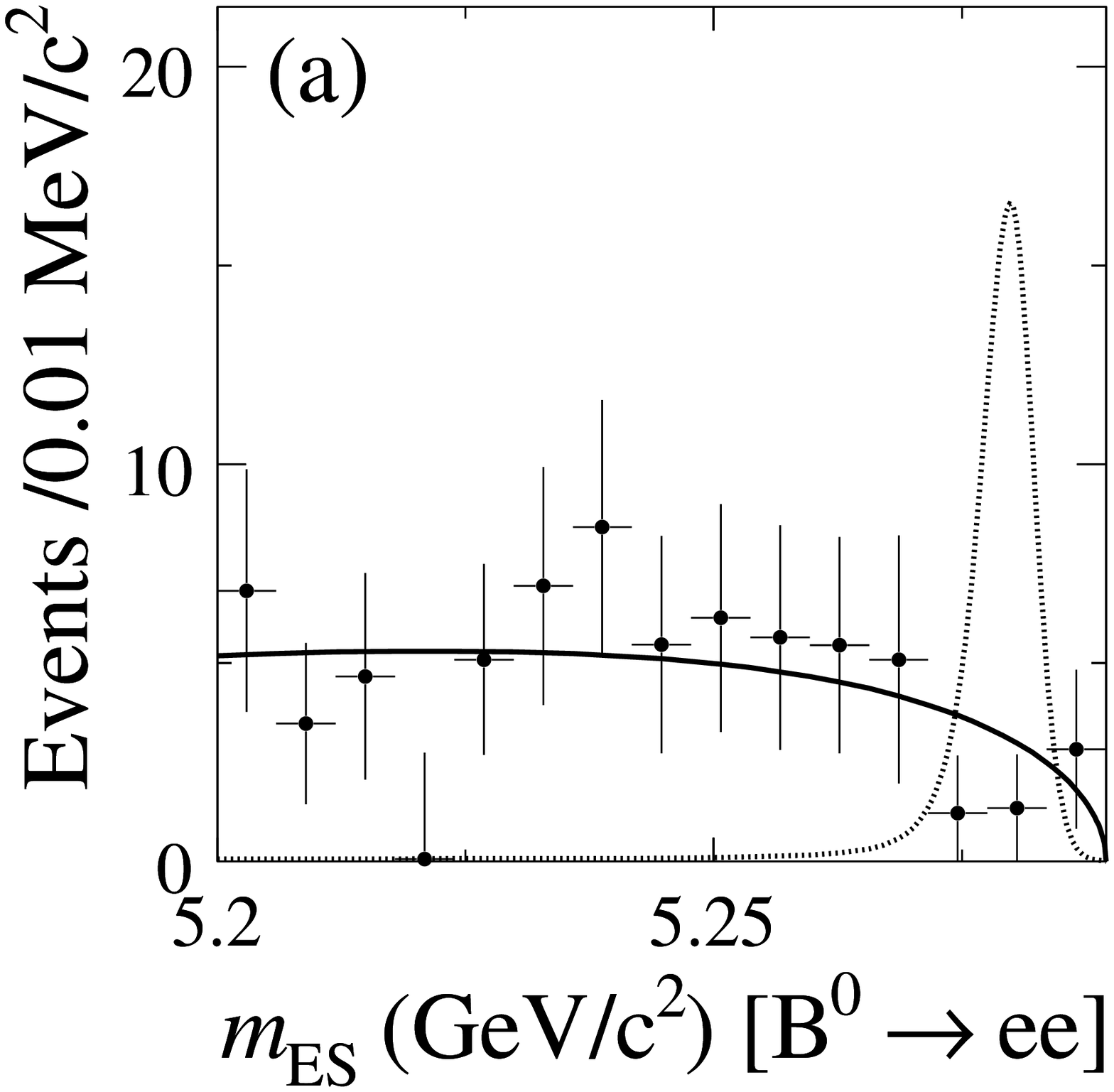}
\includegraphics[width=4.5cm]{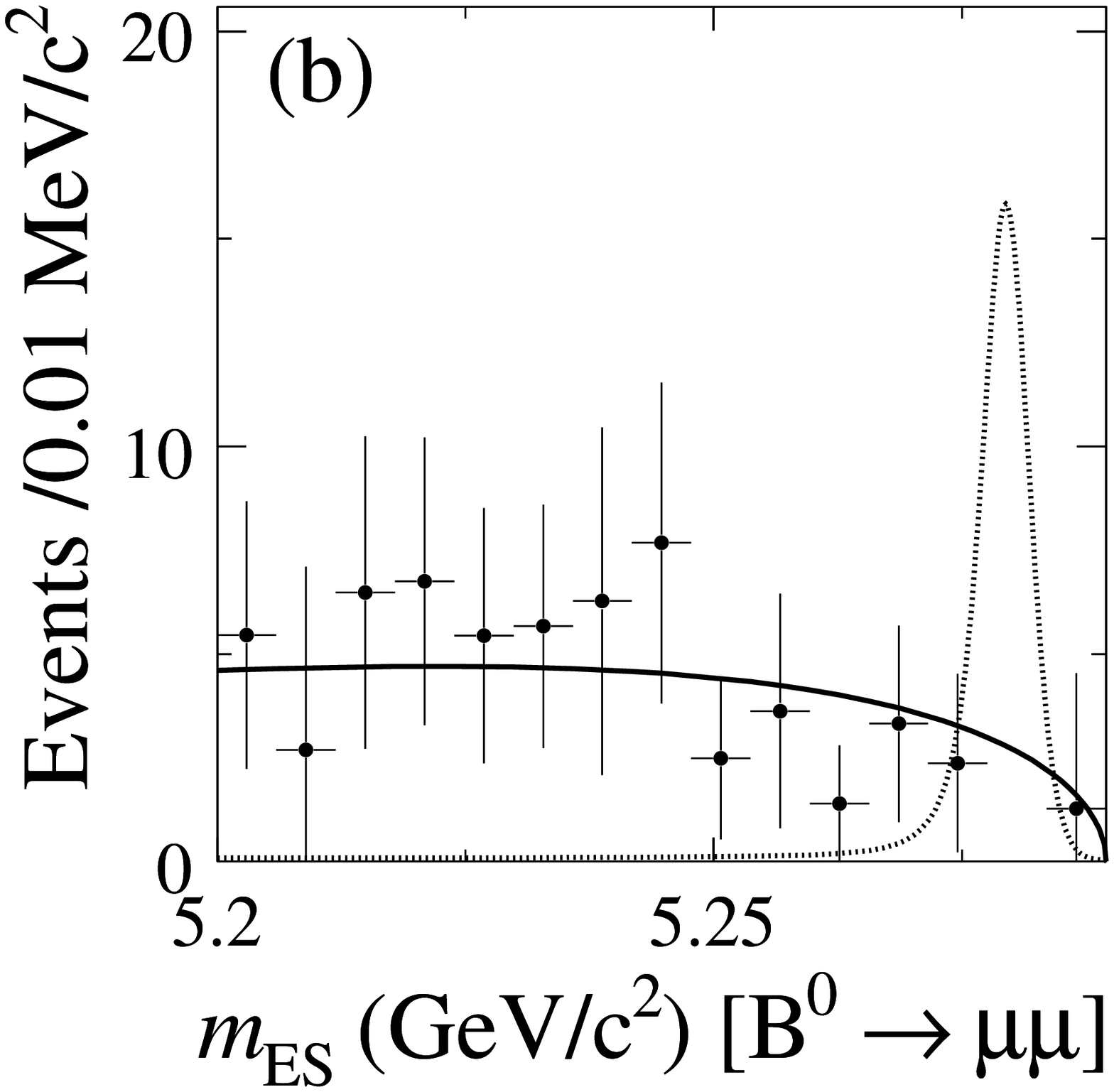}
\includegraphics[width=4.5cm]{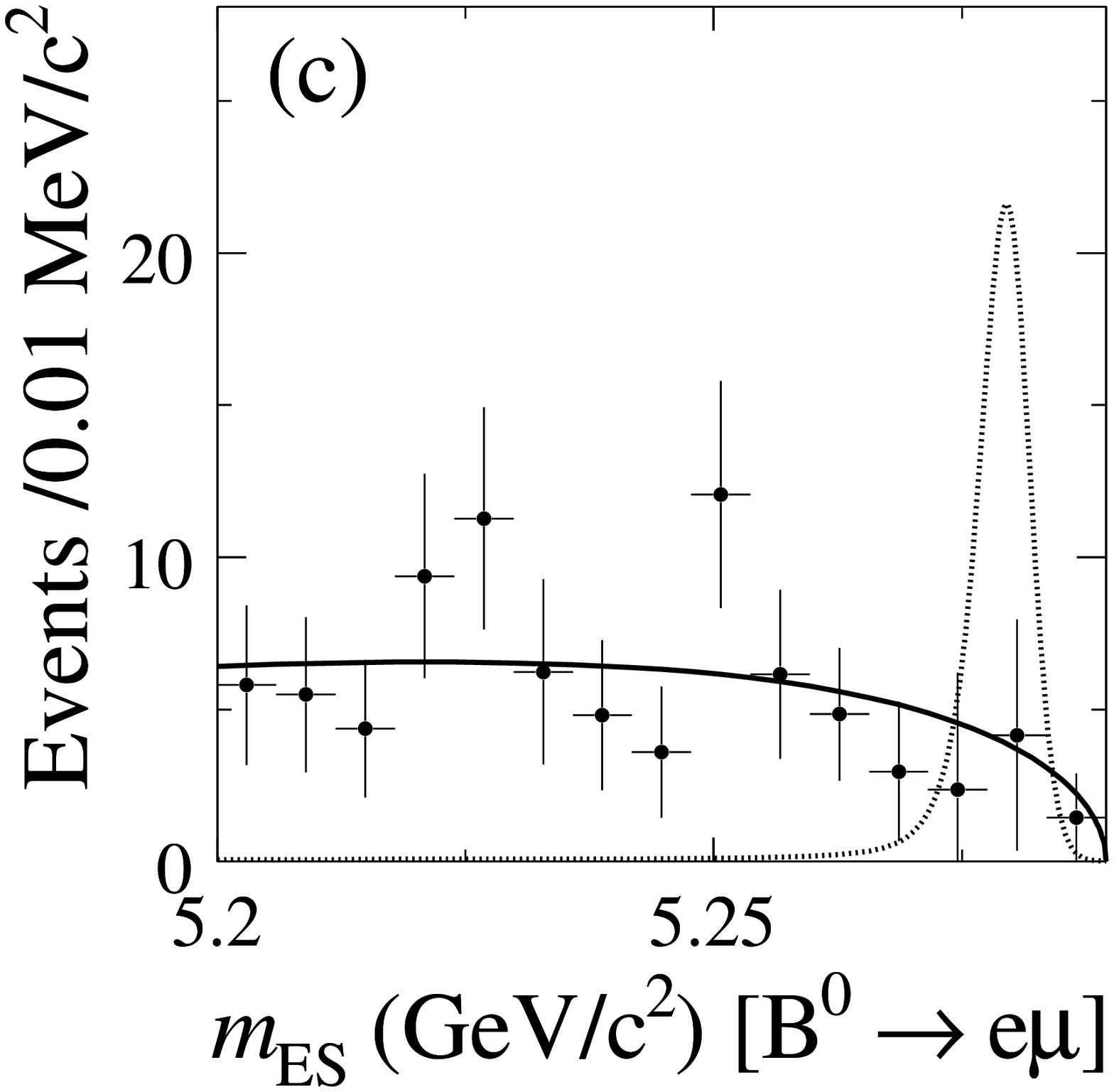}
\includegraphics[width=4.5cm]{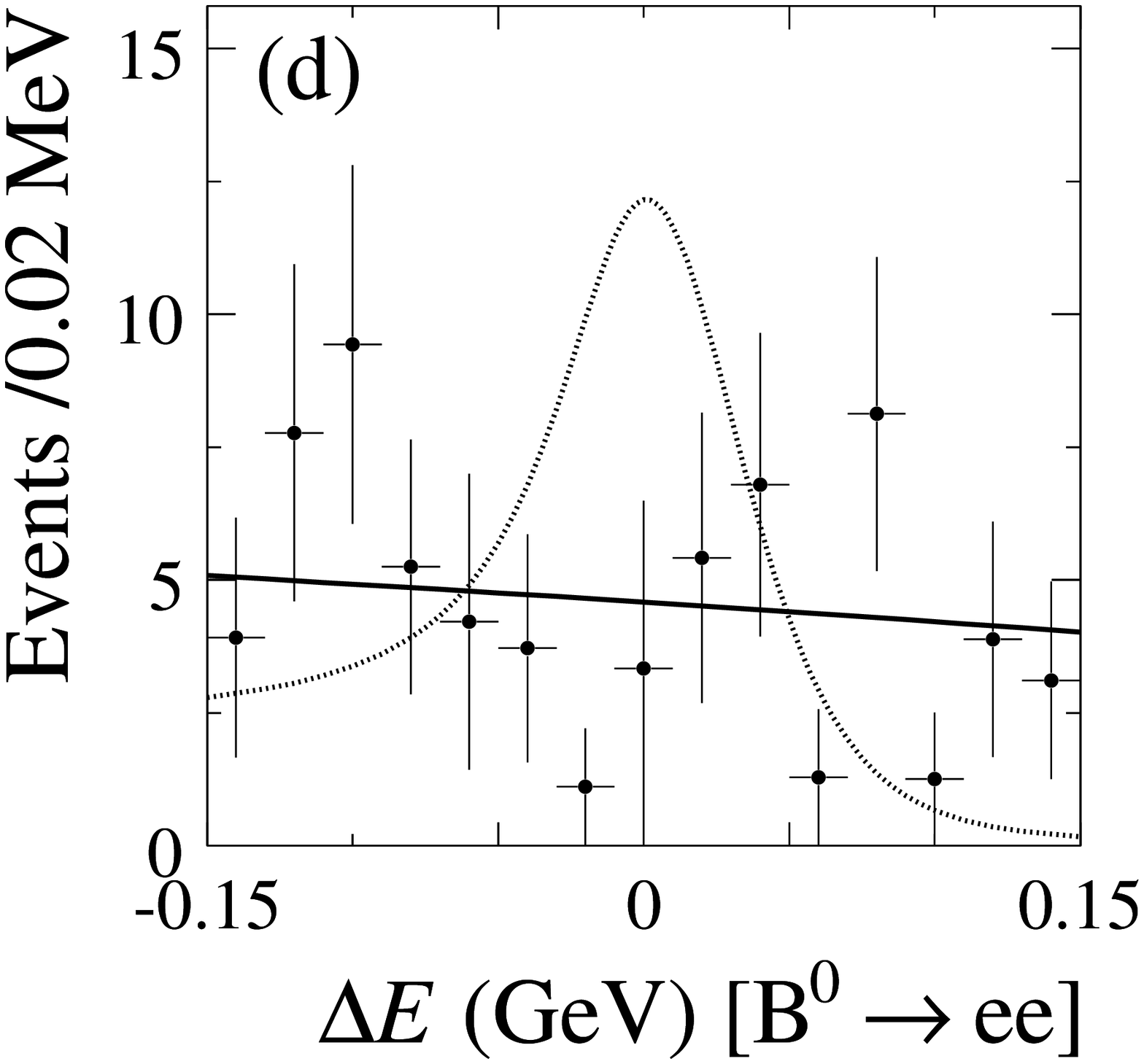}
\includegraphics[width=4.5cm]{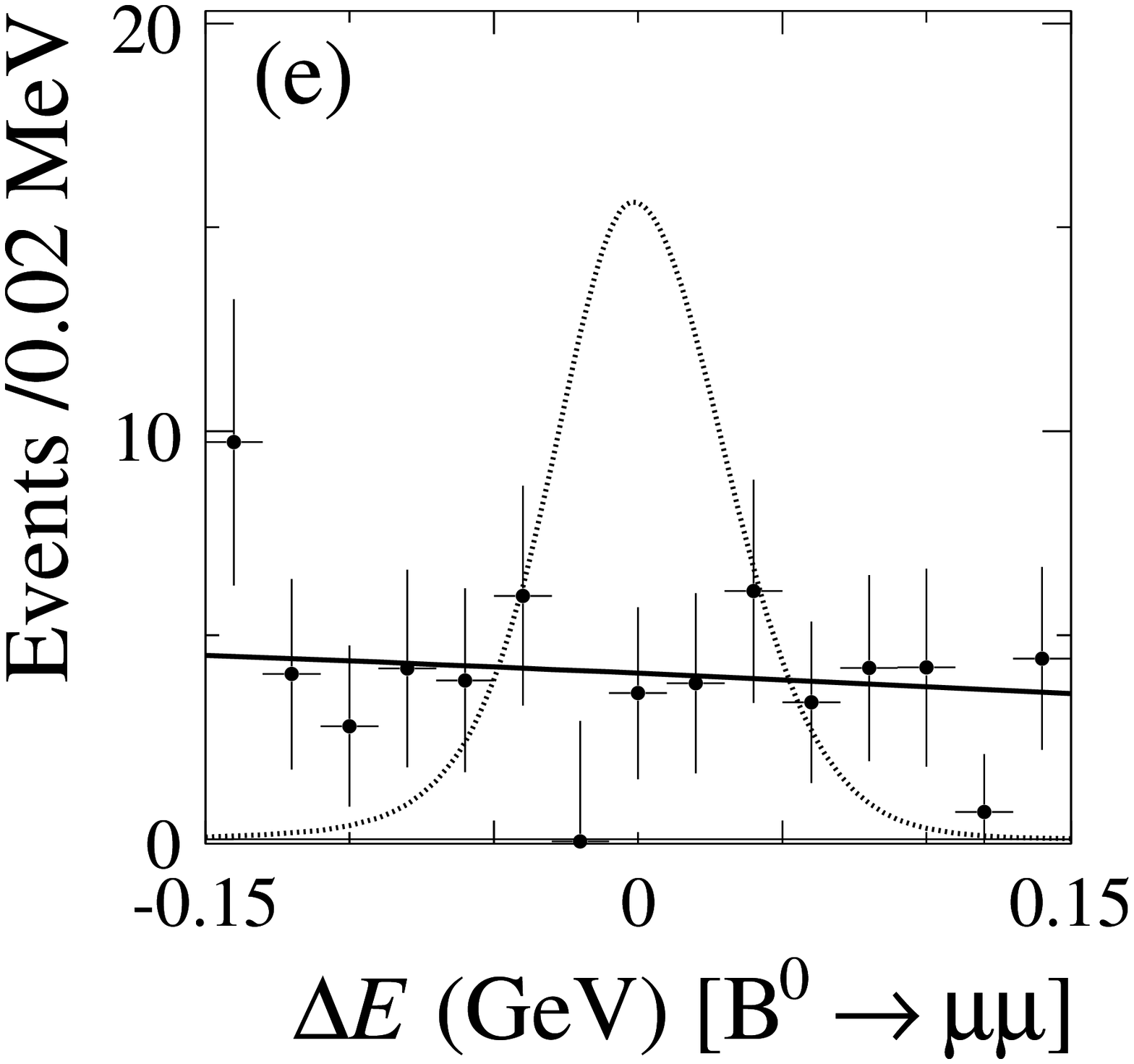}
\includegraphics[width=4.5cm]{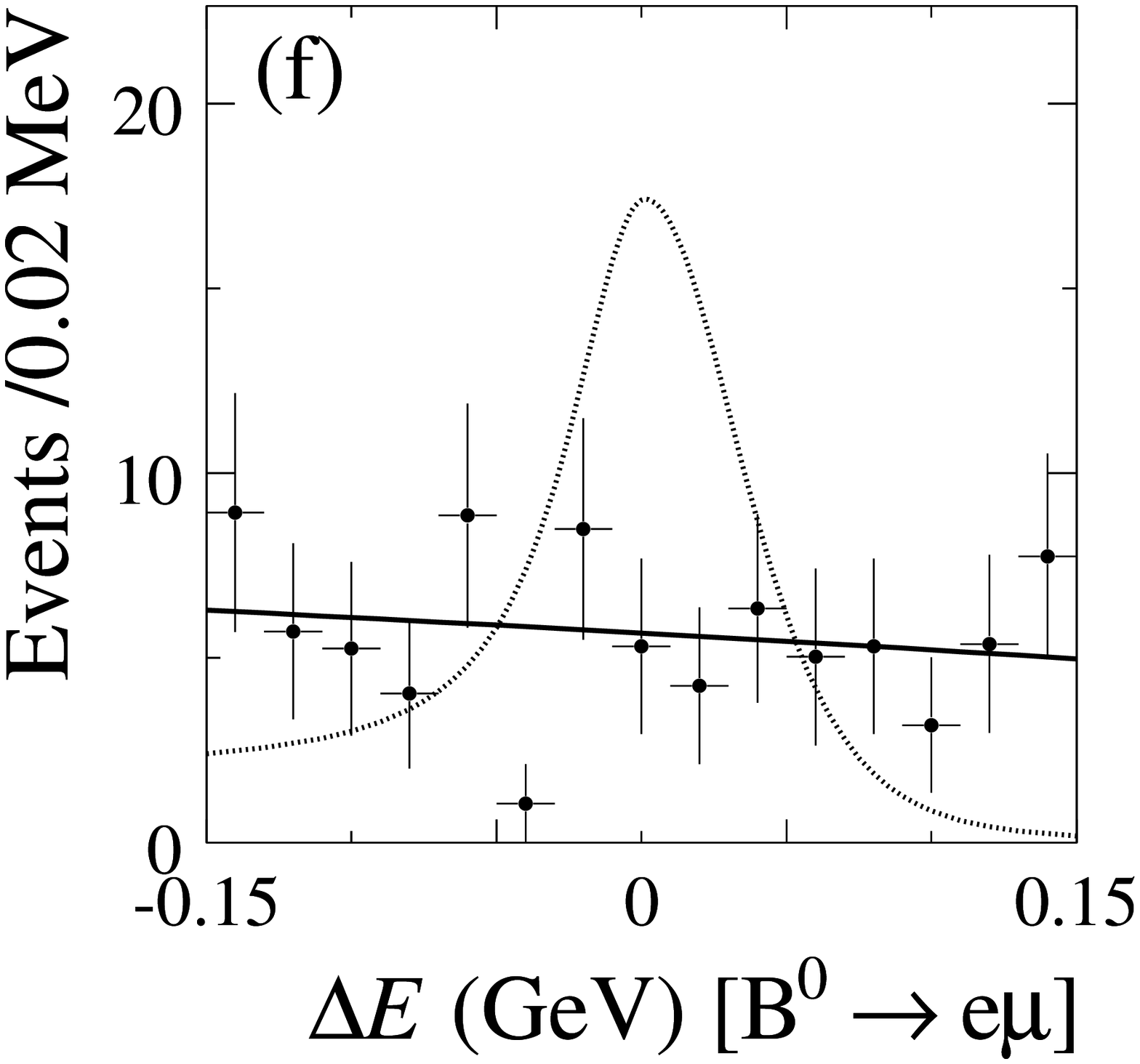}
\includegraphics[width=4.5cm]{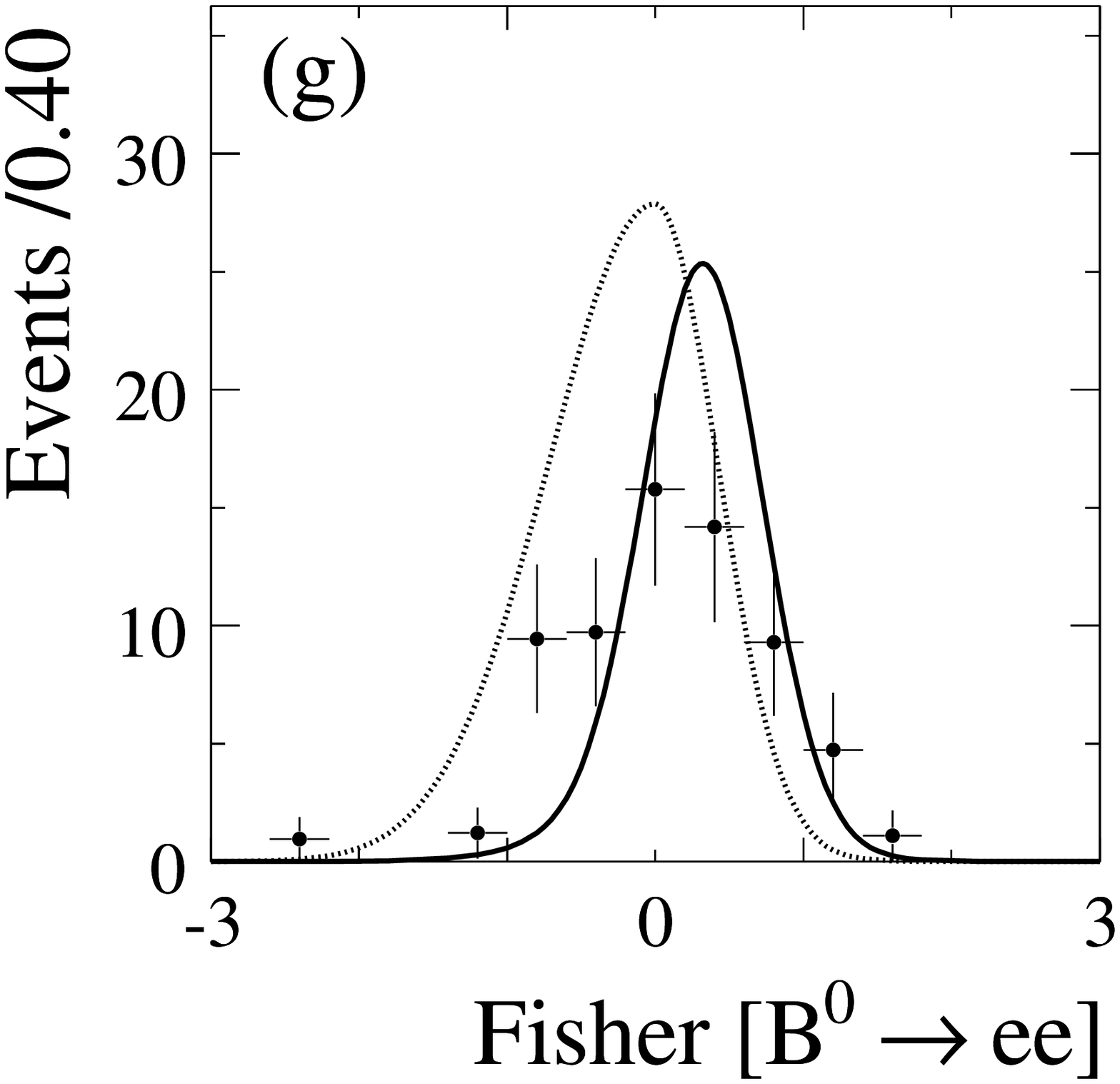}
\includegraphics[width=4.5cm]{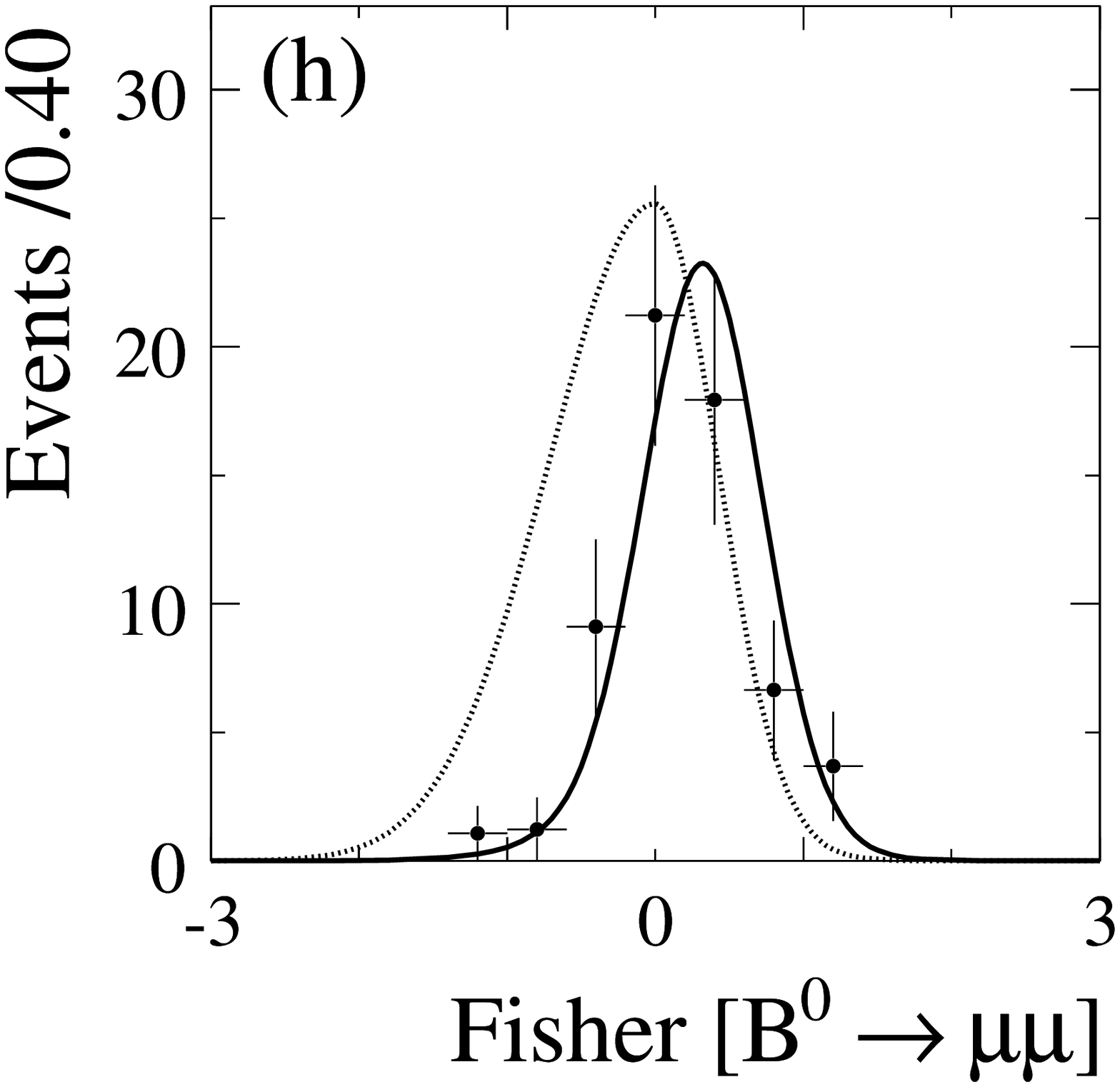}
\includegraphics[width=4.5cm]{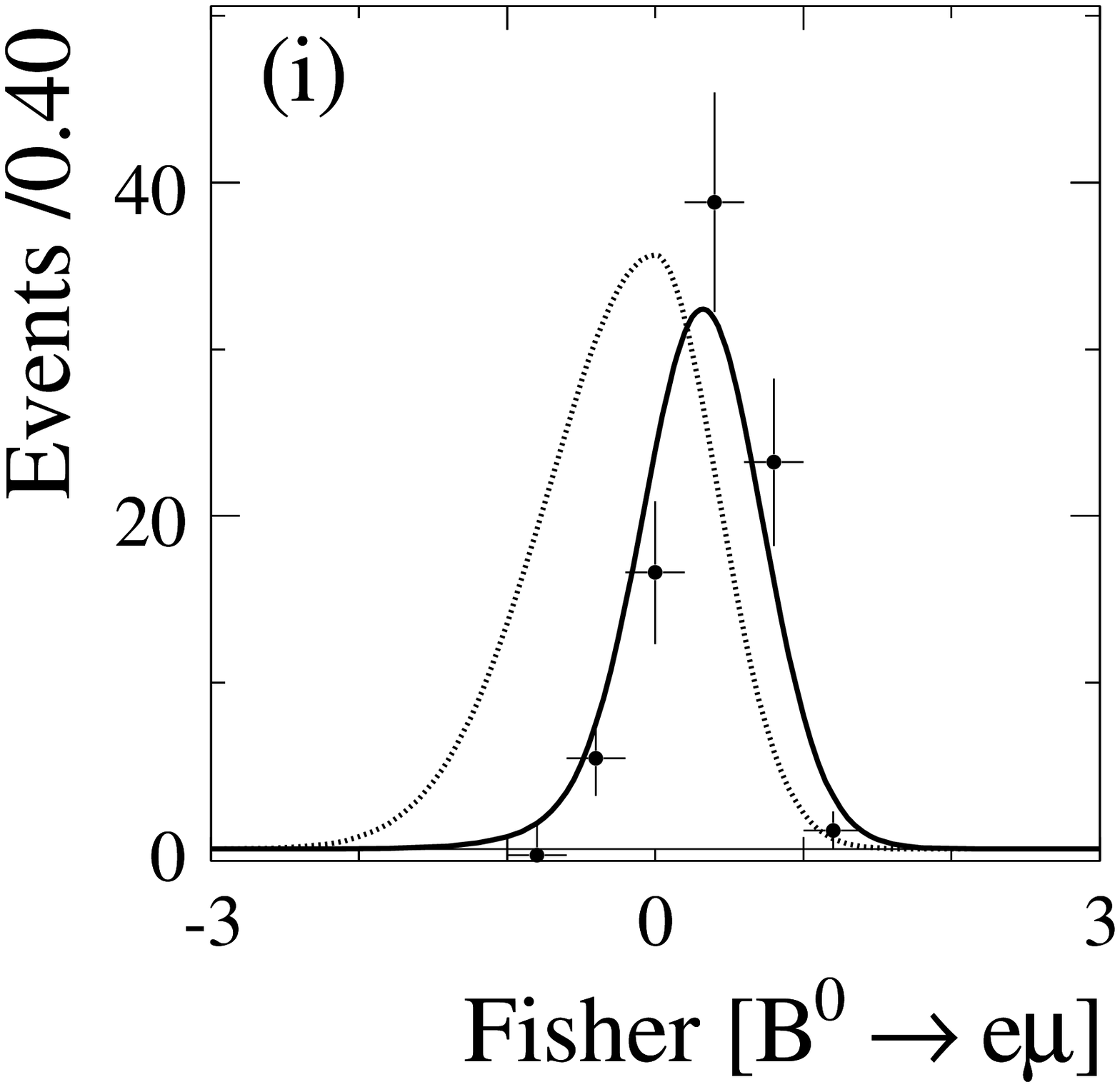}
\caption{Distribution of events in $\mes$ (a,b,c), $\de$ (d,e,f) and
  $\cal{F}$ (g,h,i) for $\bee$ (left), $\bmm$ (middle) and $\bem$
  (right).  The overlaid solid curves in each plot are the background
  $_{s}Plot$~\cite{splot} distributions obtained when the
  corresponding component is ignored in the maximum likelihood fit and
  the likelihood maximized with respect to all the other components.
  The dotted line is the PDF obtained from signal Monte Carlo with
  arbitrary normalization.}
\label{fig:result}
\end{figure*}

Among these events, the three signal yields are independently
determined by ML fits on the $2e$, $2\mu$ and $1\mu 1e$ samples.  Each
likelihood function is based on the variables \mes, $\DeltaE$ and
$\cal{F}$. The probability density functions (PDFs) for the signal
$\mes$ and $\de$ distributions are parameterized as:
\begin{equation}
f(x)=\exp \left(-\frac{(x-\mu)^2}{2\cdot \sigma_{R/L}^2 +\alpha_{R/L} \cdot
(x-\mu)^2}\right),
\end{equation}
where $\mu$ is the maximum, $\sigma_{R/L}$ represent the standard
deviation of the Gaussian component and $\alpha_{R/L}$ describe the
non-Gaussian tails of the PDF for $x > \mu$ (R) and $x < \mu$ (L).
The $\cal{F}$ distribution for signal events is described by a
Gaussian function with different RMS on the left and right side.  The
PDF of the background \mes distribution is parameterized by an
ARGUS~\cite{argus} function, the background \de~distribution by a
second order polynomial and the background $\cal{F}$~distribution by
the sum of two Gaussian functions.  Figure~\ref{fig:result} shows the
estimated background distributions for the three subsamples (solid
lines) and, just for comparison, the corresponding signal PDFs
obtained from Monte Carlo (dotted lines) with arbitrary normalization.

We find that the residual background distributions of \mes, \de~and
$\cal F$ are the same in the three leptonic samples. This has been
verified using data in the off-resonance sample and on-resonance
events populating the kinematic sidebands ($\mes< 5.27$ ~GeV/c$^2$ or
$|\DeltaE| > 150$ MeV).

In the fit the shape parameters for the \bll\ (signal) PDFs are
obtained from the MC simulation with a correction factor that accounts
for differences between data and MC, while the background PDF shape
parameters are determined on data with a procedure described below.

We determine the parameters of the background PDFs by fitting their
distribution on the $h^{+}h^{\prime -}$ sample, where we use the
Cherenkov angle to separate $B^0\to\pipi$ and $B^0\to
K^{\pm}\pi^{\mp}$.

The yields of $B^0\to\pipi$ and $K^{\pm}\pi^{\mp}$ in our
$h^{+}h^{\prime -}$ sample are consistent with the results of the
previous \babar\ analysis~\cite{pipinew}. We find $\sim 600$ signal
and $\sim 3.5\times 10^4$ background events for \pipi, $\sim 2200$
signal and $\sim 2.3\times 10^4$ background events for
$K^{\pm}\pi^{\mp}$.

The background shape parameters in the \bll\ fit are fixed to the
central values obtained in the fit to $B^0\to\pipi$ and $B^0\to
K^{\pm}\pi^{\mp}$ samples, and their errors are used to estimate the
associated systematic uncertainty on the leptonic yields.

We find no bias in the background shape parameters determined by the
procedure described above on a large number of MC simulated
$h^{+}h^{\prime -}$ event samples.

We correct for discrepancies between data and MC in the \bll\ signal
shape parameters by rescaling the PDF parameters obtained from the
simulation by the ratio between the values of the $B^0\to \pipi$ PDF
parameters in data and MC.

The knowledge of the rescaled shapes is limited by the size of the
$B^0 \to \pipi$ component in data, which causes a strong correlation
among the parameters of each signal PDF.  In order to avoid
double-counting of these effects, we take the largest observed
deviation as the systematic error induced on the leptonic yields.  The
errors on the signal yields due to the PDF shapes are $\sim 1.1$,
$\sim 0.4$ and $\sim 0.2$ events for the $e^{+}e^{-}$,
$\mu^{+}\mu^{-}$ and $e^{\pm} \mu^{\mp}$ channels, respectively.

Our results are summarized in Table~\ref{tab:result}.  We find no
evidence of signal in any of the three modes.  Using a bayesian
approach, a $90\%$ probability upper limit (UL) on the branching
fraction (\Bf ) is calculated as
\begin{equation}
\int_0^{UL}\mathcal{L}(\Bf) \,d\Bf {\bigg/} \int_0^{\infty}\mathcal{L}(\Bf) \,d\Bf = 0.9 .
\end{equation}
The \Bf\ is calculated as
\begin{equation}
\Bf \equiv \frac{N_{ll'}}{\epsilon_{ll'} N_{\BB}} \mbox{ ,}
\label{eq:br}
\end{equation}
with $N_{ll'}$ indicating the signal yield, $\epsilon_{ll'}$ the
reconstruction efficiency, and $N_{\BB}$ the number of \BB pairs in
the dataset, $N_{\BB} = (383.6\pm4.2)\times 10^6$.  We make the
assumption that the \FourS branching fractions to $\BpBm$ and $\BzBzb$
are equal.

\begin{figure}[hbtp]
\includegraphics[width=5.0cm]{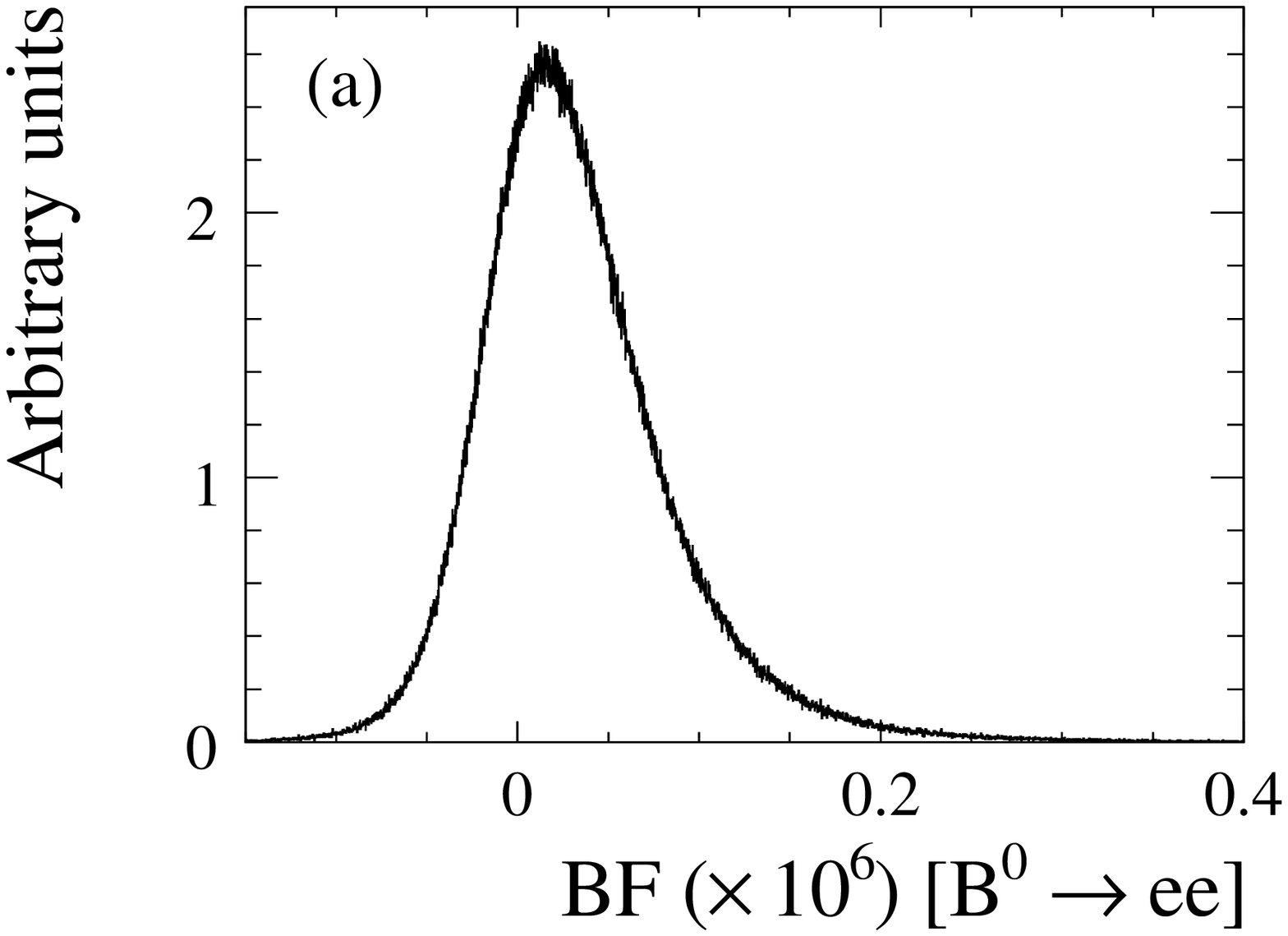} \\ 
\includegraphics[width=5.0cm]{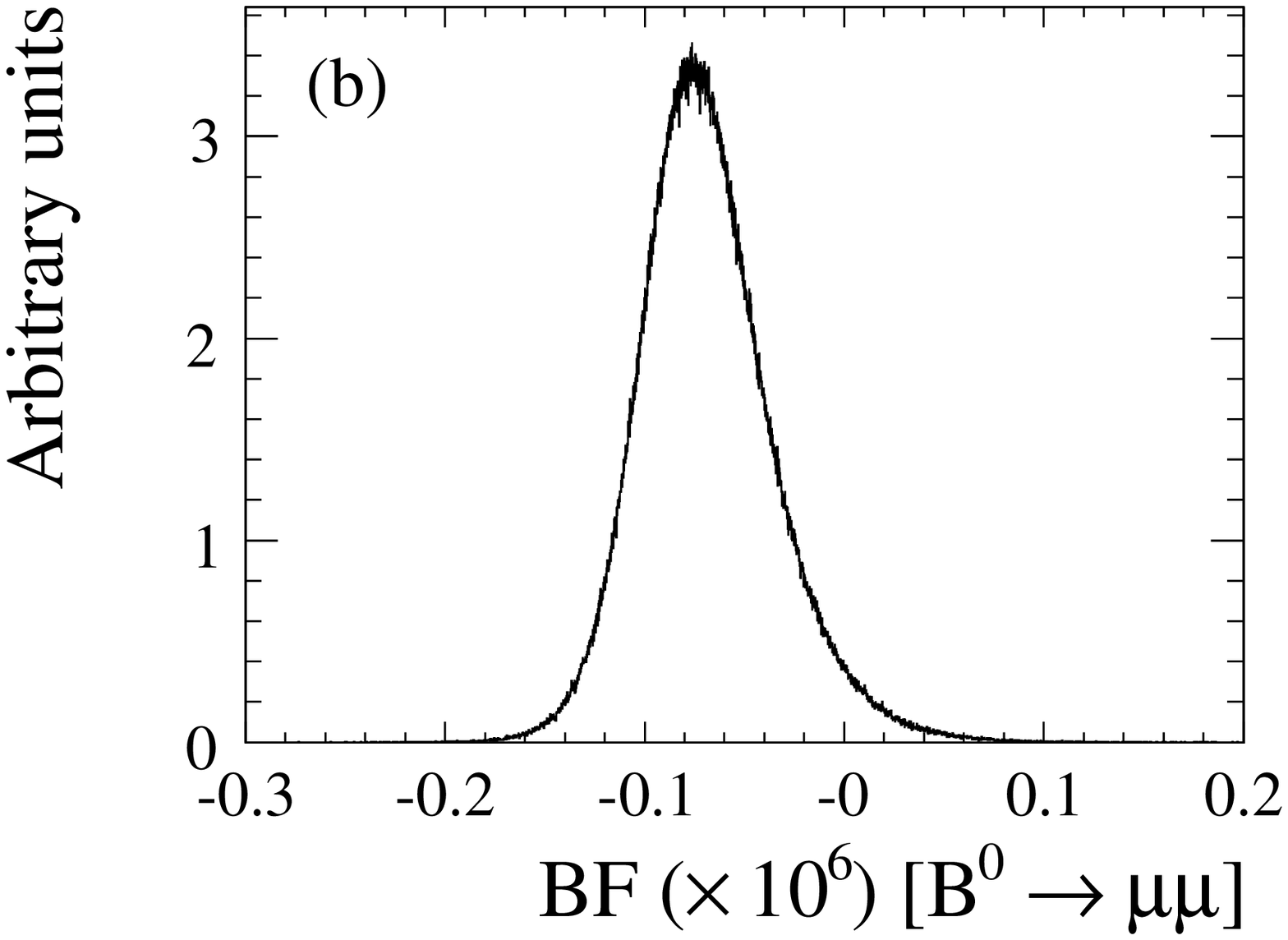} \\
\includegraphics[width=5.0cm]{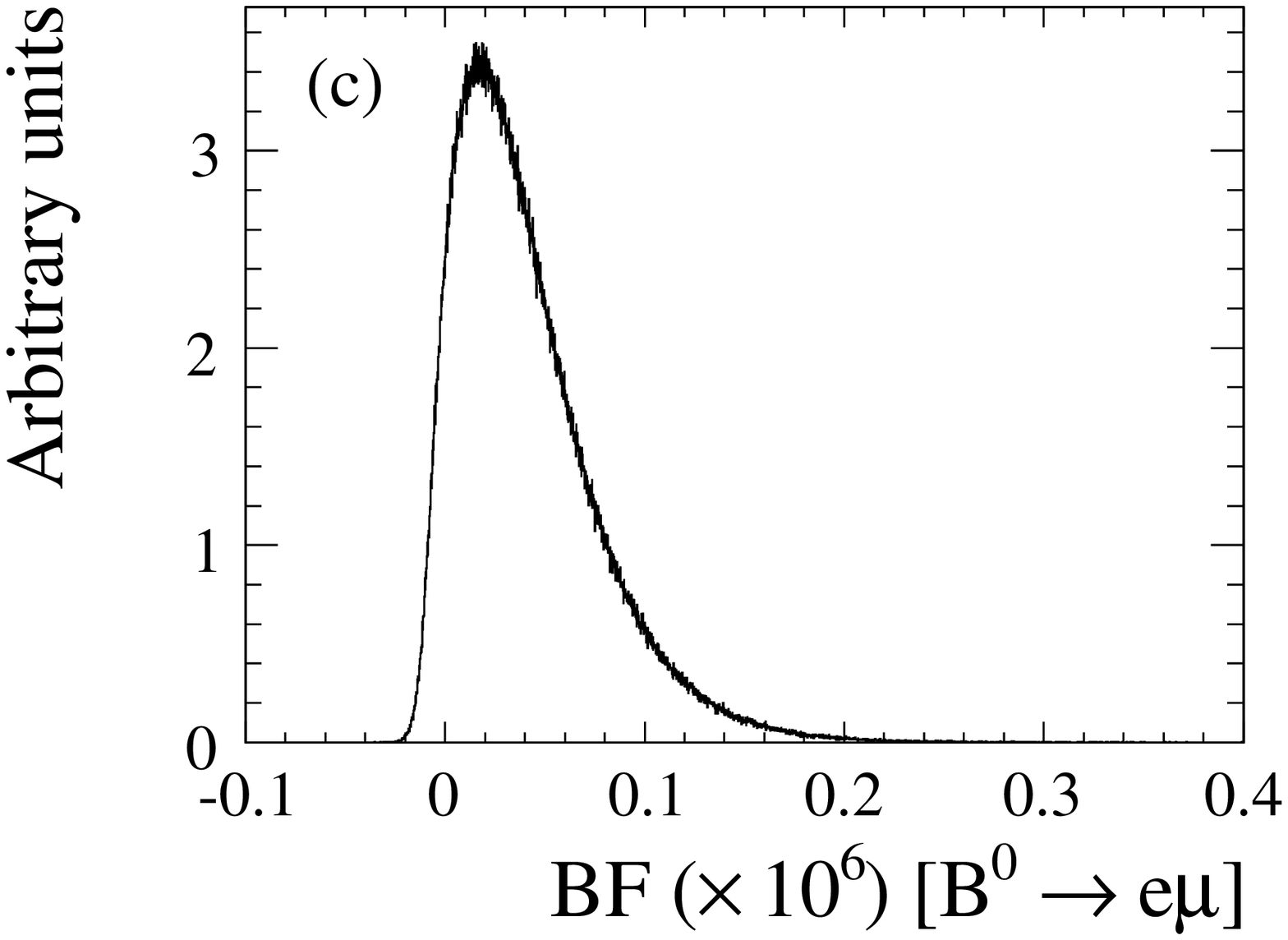}
\caption{Distribution of the likelihood as function of the \Bf\ for
\epem (a), \mumu (b) and $e^\pm\mu^\mp$ (c) decays.}
\label{fig:likelihoods}
\end{figure}

The likelihood $\cal{L}$(\Bf) is obtained by including in the
likelihood function for the signal yield ${\cal L}(N_{ll'})$ the
systematic errors on $N_{ll'}$ and the total number of \BB pairs, and
the statistical and systematic errors on the efficiency
$\epsilon_{ll'}$.  We use the relation of Eq.(\ref{eq:br}) and assume
Gaussian shapes for the errors. Figure~\ref{fig:likelihoods} shows the
likelihood distributions of the three leptonic decays.

We evaluate the efficiencies for individual selection criteria from MC
simulation and correct the results for small differences between the
simulation and the data. We take these observed differences as a
measure of the systematic uncertainties on the efficiencies.

The efficiency of PID requirements is calculated by using MC
simulations of signal events. It is then corrected with efficiency
ratios computed on data and MC, as function of track charge, momentum,
and polar angle.  We take into account the systematic error associated
to this correction.

The total systematic error on the efficiencies is $\sim 4 \%$,
calculated as the sum in quadrature of all these contributions.

In summary, we find no evidence of signal for $\bll$ and place $90\%$
confidence level upper limits on the branching fractions of
\bee , $\mu^+\mu^-$ and $e^{\pm}\mu^{\mp}$.

Table~\ref{tab:result} reports the efficiency, the number of signal
events and the UL expected in each of these modes based on MC
simulation for a sample of the size of our data sample.

The present result on $\bem$ and $\bmm$ improve the previous \babar\
upper limits~\cite{babar} based on $111~\invfb$.

The upper limit reported here for \bee\ is higher than the value
obtained in~\cite{babar}. In our previous paper we used a largely
frequentist approach~\cite{barlow} that does not explicitly require a
non-negative signal. The present results supercede our previous
results: the analysis has a higher sensitivity, estimated from the
value of the expected UL, and is based on a larger dataset that
includes the sample used in~\cite{babar}.

\begin{table}[htbp]
\begin{center}
\caption{Efficiency ($\epsilon_{ll'}$), number of signal events
($N_{ll'}$), 90\% C.L. Upper Limit on the \Bf (UL(\Bf)) for the three
leptonic decays $\bee$, $\bmm$ and $\bem$ and the corresponding
expected value based on MC simulation.}
\vspace{0.1in}
\begin{small}
\begin{tabular}{lcccc} \hline\hline\noalign{\vskip1pt} 
&$\epsilon_{ll'}(\%)$ &$N_{ll'}$ &$\mbox{UL(\Bf)}\times 10^{-8}$	&Exp(UL)   \\
\hline\noalign{\vskip1pt}
$B^0 \to e^+e^-$       &$16.6\pm0.3$        &$0.6\pm2.1$  & $11.3$	&$7.4$	\\
$B^0 \to \mu^+\mu^-$   &$15.7\pm0.2$        &$-4.9\pm1.4$ & $5.2$	&$5.9$	\\
$B^0 \to e^\pm \mu^\mp$ &$17.1\pm0.2$        &$1.1\pm1.8$  & $9.2$	&$6.3$	\\
\hline\hline\noalign{\vskip1pt}
\end{tabular}
\end{small}
\label{tab:result}
\end{center}
\end{table}

\input pubboard/acknow_PRL.tex

\end{document}

%% file: pubboard/authors_jul2007.tex
%
\author{B.~Aubert}
\author{M.~Bona}
\author{D.~Boutigny}
\author{Y.~Karyotakis}
\author{J.~P.~Lees}
\author{V.~Poireau}
\author{X.~Prudent}
\author{V.~Tisserand}
\author{A.~Zghiche}
\affiliation{Laboratoire de Physique des Particules, IN2P3/CNRS et Universit\'e de Savoie, F-74941 Annecy-Le-Vieux, France }
\author{J.~Garra~Tico}
\author{E.~Grauges}
\affiliation{Universitat de Barcelona, Facultat de Fisica, Departament ECM, E-08028 Barcelona, Spain }
% ...
\collaboration{The \babar\ Collaboration}
\noaffiliation

%% file: pubboard/acknow_PRL.tex
We are grateful for the excellent luminosity and machine conditions
provided by our \pep2\ colleagues, 
and for the substantial dedicated effort from
the computing organizations that support \babar.
The collaborating institutions wish to thank 
SLAC for its support and kind hospitality. 
This work is supported by
DOE
and NSF (USA),
NSERC (Canada),
CEA and
CNRS-IN2P3
(France),
BMBF and DFG
(Germany),
INFN (Italy),
FOM (The Netherlands),
NFR (Norway),
MES (Russia),
MEC (Spain), and
STFC (United Kingdom). 
Individuals have received support from the
Marie Curie EIF (European Union) and
the A.~P.~Sloan Foundation.

%% file: main.bbl
\begin{thebibliography}{99}

\bibitem{susy} 
  K.~S.~Babu and C.~F.~Kolda,
  Phys.\ Rev.\ Lett.\  {\bf 84}, 228 (2000);
  P.~H.~Chankowski and L.~Slawianowska,
  Phys.\ Rev.\  D {\bf 63}, 054012 (2001);
  C.~Bobeth {\em et al.},
  Phys.\ Rev.\  D {\bf 64}, 074014 (2001).

\bibitem{BRSM}
  M.~Misiak and J.~Urban,
  Phys.\ Lett.\  B {\bf 451}, 161 (1999);
  G.~Buchalla and A.~J.~Buras,
  Nucl.\ Phys.\  B {\bf 548}, 309 (1999).

\bibitem{ref:tautau}
   \babar\ Collaboration, B.\ Aubert {\em et al.},
   Phys.\ Rev.\ D {\bf 96}, 241802 (2006).

\bibitem{burasemu}
  M.~Blanke {\em et al.},
  JHEP {\bf 0705}, 013 (2007);
  R.~A.~Diaz,
  Eur.\ Phys.\ J.\ C {\bf 41}, 305 (2005);
  A.~Ilakovac
  Phys.\ Rev.\ D {\bf 62}, 036010 (2000).

\bibitem{MFV}
  A.~J.~Buras,
  Acta\ Phys.\ Polon.\ B {\bf 34}, 5615 (2003);
  G.~D'Ambrosio {\em et al.},
  Nucl.\ Phys.\  B {\bf 645}, 155 (2002).

\bibitem{UUT}
  A.~J.~Buras {\em et al.},
  Phys.\ Lett.\ B {\bf 500}, 161 (2001).

\bibitem{bobeth}
  C.~Bobeth {\em et al.},
  Nucl.\ Phys.\  B {\bf 726}, 252 (2005).

\bibitem{gino}
  G.~Isidori and A.~Retico,
  JHEP {\bf 0111}, 001 (2001);
  G.~Isidori and P.~Paradisi,
  Phys.\ Lett.\  B {\bf 639}, 499 (2006).

\bibitem{babar} \babar\ Collaboration, B.\ Aubert {\em et al.},
Phys. Rev. Lett. {\bf 94}, {221803} (2005).

\bibitem{belle03} Belle Collaboration, M.-C. Chang {\em et al.}, Phys. Rev. D~{\bf 68},
 {111101(R)} (2003).

\bibitem{previousCDF}{
 	CDF Public note $\#8176$.
    	{\url{http://www-cdf.fnal.gov/physics/new/bottom/060316.blessed-bsmumu3/bsmumupub3_v02.ps}}
	}.

\bibitem{cleo} CLEO Collaboration, T. Bergfeld {\em et al.}, Phys. Rev. D~{\bf 62},
 {091102(R)} (2000).

\bibitem{babarnim}
\babar\ Collaboration, B.\ Aubert {\em et al.}, 
Nucl.\ Instrum.\ Methods~{\bf A479}, 1 (2002).

\bibitem{geant}
S.~Agostinelli {\em et al.},
Nucl.\ Instrum.\ Methods~{\bf A506}, 250 (2003).

\bibitem{sphericity} S. L. Wu, Phys. Rep.~{\bf 107}, 59 (1984).

\bibitem{foxwolf} G.C. Fox and S. Wolfram, \prl {\bf 41}, 1581 (1978). 

\bibitem{fisher} R.~A.~Fisher, Annals Eugen. {\bf 7}, 179 (1936).

\bibitem{thrust} 
  S.~Brandt {\it et al.} Phys. Lett. {\bf 12}, 57 (1964);
  E.~Farhi, Phys. Rev. Lett. {\bf 39}, 1587 (1977).

\bibitem{argus} H. ~Albrecht {\it et al.},	[ARGUS Collaboration], Phys. Lett. B {\bf 241}, 278 (1990)

\bibitem{splot} 
  M.~Pivk and F.~R.~Le Diberder,
  Nucl.\ Instrum.\ Meth.\  A {\bf 555}, 356 (2005).

\bibitem{pipinew} 
  B.~Aubert {\it et al.}  [BABAR Collaboration],
  Phys.\ Rev.\  D {\bf 75}, 012008 (2007).

\bibitem{barlow}
  R.~Barlow,
  Comput. Phys. Commun. {\bf 149}, 97 (2002).

\end{thebibliography}
